\documentclass[aps,prd,onecolumn,showpacs,nofootinbib,amsmath,amssymb,amsfonts,showkeys]{revtex4}
\usepackage{graphicx}
\usepackage{bm}
\usepackage{xcolor}

\begin{document}

\title{Estimating the number of solutions equation of $N$-point \\ gravitational lens algebraic geometry methods}
\author{A.T. Kotvytskiy}
\email{kotvytskiy@gmail.com}
\author{S.D. Bronza}
\email{bronza_semen@mail.ua}
\author{S.R. Vovk}
\affiliation{Karazin Kharkov National University, Ukraine}
\affiliation{Ukrainian State University of Railway Transport, Ukraine}

\begin{abstract}
One of the main problems in the study of system of equations of the
gravitational lens, is the computation of coordinates from the known position
of the source. In the process of computing finds the solution of equations
with two unknowns. The difficulty lies in the fact that, in general, is not
known constructive or analytical algorithm for solving systems of polynomial
equations In this connection, use numerical methods like the method of
tracing. For the $N$-point gravitational lenses have a system of polynomial
equations. Systems Research is advisable to start with an assessment of the
number of solutions. This can be done by methods of algebraic geometry.

\end{abstract}


\maketitle

\section{Introduction}

According to the general theory of relativity, the light beam, which passes
close to a point source of gravity (gravitational lens) at a distance $\xi$
from it (in case $\xi\gg r_{g}$ ) is deflected by an angle%
\begin{equation}
\vec{\alpha}=\frac{2r_{g}}{\xi^{2}}\vec{\xi}=\frac{4GM}{c^{2}\xi^{2}}\vec{\xi}
\label{Angle}%
\end{equation}

where $r_{g}$ - gravitational radius; $M$ - mass point of the lens; $G$ -
gravity constant; $c$ - velocity of light in vacuum.

The detailed derivation of the formula (\ref{Angle}) can be found in many
classic books \cite{Bliokh} - \cite{Landau}. For $N$ - point of the
gravitational lens, in the case of small tilt angles have the following
equation in dimensionless variables [4], [5]:%
\begin{equation}
\vec{y}=\vec{x}-\sum\limits_{i=1}m_{i}\frac{\vec{x}-\vec{l}_{i}}{\left\vert
\vec{x}-\vec{l}_{i}\right\vert ^{2}}, \label{2}%
\end{equation}

where $\vec{l}_{i}$- dimensionless radius vector of point masses outside the
lens, and the mass $m_{i}$ satisfy the relation $%
{\displaystyle\sum}
m_{i}=1$.

The Equation (\ref{2}) in coordinate form has the form of system:%
\begin{equation}
\left\{
\begin{array}
[c]{c}%
y_{1}=x_{1}-\sum\limits_{i=1}^{N}m_{i}\frac{x_{1}-a_{i}}{_{(x_{1}-a_{i}%
)}2_{+(x_{2}-b_{i})}2}\\
y_{2}=x_{2}-\sum\limits_{i=1}^{N}m_{i}\frac{x_{2}-b_{i}}{_{(x_{1}-a_{i}%
)}2_{+(x_{2}-b_{i})}2}%
\end{array}
\right.  \label{3}%
\end{equation}

where $a_{i}$ and $b_{i}$ are the coordinates of the radius-vector $\vec
{l}_{i}$, i.e. $\vec{l}_{i}=(a_{i},b_{i})$ .

The right parts the equations of system (\ref{3}), are rational functions of
the variables $x_{1}$ and $x_{2}$. We transform each equation of the system
(\ref{3}) in a polynomial equation, and we obtain a system of equations:%

\begin{equation}
\left\{
\begin{array}
[c]{l}%
F_{1}\left(  \,x_{1},\,x_{2},y_{1}\right)  =0\\
F_{2}\left(  \,x_{1},\,x_{2},y_{2}\right)  =0
\end{array}
\right.  \,, \label{4}%
\end{equation}

To study the solutions of the system (\ref{4}), it will be convenient methods
of algebraic geometry. \newline Indeed, the main object of study of classical
algebraic geometry, as well as in a broad sense and modern algebraic geometry,
are the set of solutions of algebraic systems, in particular polynomial,
equations. This situation gives us the opportunity to apply the techniques of
algebraic geometry in the theory of $N$-point gravitational lenses \cite{Kotv}
- \cite{Van}.

Most of the results that we need stated in terms of homogeneous coordinates to
projective curves that are defined in the projective plane.\newline Therefore,
we need, in the beginning, the system transform (\ref{4}) to a form convenient
for evaluation.\newline To transform the system (\ref{4}) we need the
following terms and definitions.

\section{Mathematical definitions and theorems}

Before that provide definitions, we describe the projective plane and
homogeneous (projective) coordinates. The real projective plane can be thought
of as the Euclidean plane with additional points added, which are called
points at infinity, and are considered to lie on a new line, the line at
infinity. There is a point at infinity corresponding to each direction
(numerically given by the slope of a line), informally defined as the limit of
a point that moves in that direction away from the origin. Parallel lines in
the Euclidean plane are said to intersect at a point at infinity corresponding
to their common direction. Given a point $(x,y)$ on the Euclidean plane, for
any non-zero real number $S$, the triple $(xS,yS,S)$ is called a set of
homogeneous coordinates for the point. By this definition, multiplying the
three homogeneous coordinates by a common, non-zero factor gives a new set of
homogeneous coordinates for the same point. In particular, $(1,2)$ is such a
system of homogeneous coordinates for the point $(x,y)$. For example, the
Cartesian point $(x,y,1)$can be represented in homogeneous coordinates as
$(1,2.1)$ or $(2,4.2)$. The original Cartesian coordinates are recovered by
dividing the first two positions by the third. Thus unlike Cartesian
coordinates, a single point can be represented by infinitely many homogeneous coordinates.

Some authors use different notations for homogeneous coordinates which help
distinguish them from Cartesian coordinates. The use of colons instead of
commas, for example $(x:y:z)$ instead of $(x,y,z)$, emphasizes that the
coordinates are to be considered ratios \cite{Reid}. Square brackets, as in
$[x,y,z]$ emphasize that multiple sets of coordinates are associated with a
single point \cite{Garner}. Some authors use a combination of colons and
square brackets, as in $[x:y:z]$ \cite{Bocher}.

\subsection{The properties of homogeneous coordinates on the plane}

We define homogeneous projective coordinates for the first points of the
projective plane not lying on a straight $\infty$.\newline At all points of
the projective plane, in addition to lying on a straight $\infty$ (the line at
infinity) are homogeneous coordinates of the projective: three numbers, not
both zero.

The basic points for arithmetization projective plane (i.e., the introduction
of non-homogeneous projective coordinates) are the origin Systems; $\infty
_{x}$ (infinity on the $x$-axis), $\infty_{y}$ (infinity on the $y$-axis),
$(1,1)$- unit. Obviously, the line at infinity passes through the points
$\infty_{x}$ and $\infty_{y}$.

We give a precise definition.\newline Homogeneous coordinates of a point M is
said to be three numbers $X_{1},X_{2},X_{3}$, not both zero and such that
$\frac{X_{1}}{X_{3}}=x$ ; $\frac{X_{2}}{X_{3}}=y$, where $x$ and $y$ -
projective heterogeneous (affine) position. Homogeneous coordinates of points
$M_{\infty}$ lying on a line , $\infty$ , call three numbers $X_{1}$ , $X_{2}$
, $X_{3,}$ under conditions:

1. $X_{3}=0$;\newline2. Of two numbers $X_{1},X_{2}$ is at least one
non-zero;\newline3. The ratio $\frac{X_{1}}{X_{2}}$is equal to $\frac{B}%
{(-A)}$, where $A$ and $B$ - coefficients of each line $Ax+By+C=0$, passing
through $M_{\infty}$.

Let us look at some of the properties of homogeneous coordinates in the
projective plane.\newline1) Each point of the projective plane is the
homogeneous coordinates\newline2) If $X_{1},X_{2},X_{3}$ - homogeneous
coordinates of the point $M$, then $sX_{1},sX_{2},sX_{3}$ (where $s$ - any
non-zero number), too, are homogeneous coordinates of the point $M$.\newline3)
different points correspond to different attitudes $\frac{X_{1}}{X_{3}}$;
$\frac{X_{2}}{X_{3}}$their homogeneous coordinates.\newline4) Direct
$A_{1}A_{2}$ - is the line at infinity - it is in homogeneous coordinates the
equation $X_{3}=0$.\newline5) The axes have their usual equation.

6) If the equation of the line $Ax+By+C=0$ substitute homogeneous coordinates
of a point $M$, lying on the straight line ( $\frac{X_{1}}{X_{3}}=x$%
;$\frac{X_{2}}{X_{3}}=y$), then we get: $AX_{1}+BX_{2}+CX_{3}=0$, linear
equation in a homogeneous form (there is no free member)\newline7) The
equation of any curve in homogeneous coordinates is a homogeneous equation,
and its degree is called the degree of the curve.

8) A polynomial $g(x,y)$ of degree $k$ can be turned into a homogeneous
polynomial by replacing $x$ with $\frac{X_{1}}{X_{3}}$, $y$ with $\frac{X_{2}%
}{X_{3}}$, and multiplying by $X_{3}^{k}$, in other words by defining.

The resulting function $P$ is a polynomial so it makes sense to extend its
domain to triples where $X_{3}=0$. The equation $P\left(  X_{1},X_{2}%
,X_{3}\right)  =0$

can then be thought of as the homogeneous form of $g(x,y)$, and it defines the
same curve when restricted to the Euclidean plane.

Transform the system (\ref{3}) into the equation system in homogeneous
coordinates. We introduce the notation of homogeneous coordinates.

Let $x_{1}=\frac{X_{1}}{X_{0}},\;\;x_{2}=\frac{X_{2}}{X_{0}}$. We have a
system of equations in homogeneous coordinates:%

\begin{equation}
\left\{
\begin{array}
[c]{l}%
y_{1}=\frac{X_{1}}{X_{0}}-\sum\limits_{i=1}^{N}m_{i}\frac{\frac{X_{1}}{X_{0}%
}-a_{i}}{\left(  \frac{X_{1}}{X_{0}}-a_{i}\right)  ^{2}+\left(  \frac{X_{2}%
}{X_{0}}-b_{i}\right)  ^{2}}\\
y_{2}=\frac{X_{2}}{X_{0}}-\sum\limits_{i=1}^{N}m_{i}\frac{x_{2}-b_{i}}{\left(
\frac{X_{1}}{X_{0}}-a_{i}\right)  ^{2}+\left(  \frac{X_{2}}{X_{0}}%
-b_{i}\right)  ^{2}}%
\end{array}
\right.  \label{5}%
\end{equation}

We transform the system (\ref{5})%

\begin{equation}
\left\{
\begin{array}
[c]{l}%
y_{1}=\frac{X_{1}}{X_{0}}-\sum\limits_{i=1}^{N}m_{i}\frac{\left(  X_{1}%
-a_{i}X_{0}\right)  X_{0}}{\left(  X_{1}-a_{i}X_{0}\right)  ^{2}+\left(
X_{2}-b_{i}X_{0}\right)  ^{2}}\\
y_{2}=\frac{X_{2}}{X_{0}}-\sum\limits_{i=1}^{N}m_{i}\frac{\left(  X_{2}%
-b_{i}X_{0}\right)  X_{0}}{\left(  X_{1}-a_{i}X_{0}\right)  ^{2}+\left(
X_{2}-b_{i}X_{0}\right)  ^{2}}%
\end{array}
\right.  \label{6}%
\end{equation}

We transform the system (\ref{6}) to form:%

\begin{equation}
\left\{
\begin{array}
[c]{l}%
F_{1}\left(  \,X_{0},X_{1},X_{2},y_{1}\right)  =0\\
F_{2}\left(  \,X_{0},X_{1},X_{2},y_{2}\right)  =0
\end{array}
\right.  \qquad\, \label{7}%
\end{equation}

where $F_{1}=$ $F_{1}\left(  \,X_{0},X_{1},X_{2},y1\right)  $ and $F_{2}=$
$F_{2}\left(  \,X_{0},X_{1},X_{2},y_{2}\right)  $- homogeneous polynomials.

In equations of system (\ref{6}), under the sign of the amount we give to a
common denominator. We have:%

\begin{equation}
\left\{
\begin{array}
[c]{l}%
y_{1}=\frac{X_{1}}{X_{0}}-\frac{X_{0}\sum\limits_{j=1}^{N}m_{j}\left(
X_{1}-a_{j}X_{0}\right)  \left\{
{\displaystyle\prod\limits_{i=1,i\neq j}^{N}}
\left[  \left(  X_{1}-a_{i}X_{0}\right)  ^{2}+\left(  X_{2}-b_{i}X_{0}\right)
^{2}\right]  \right\}  }{%
{\displaystyle\prod\limits_{i=1}^{N}}
\left[  \left(  X_{1}-a_{i}X_{0}\right)  ^{2}+\left(  X_{2}-b_{i}X_{0}\right)
^{2}\right]  }\\
y_{2}=\frac{X_{2}}{X_{0}}-\frac{X_{0}\sum\limits_{j=1}^{N}m_{j}\left(
X_{2}-b_{j}X_{0}\right)  \left\{
{\displaystyle\prod\limits_{i=1,i\neq j}^{N}}
\left[  \left(  X_{1}-a_{i}X_{0}\right)  ^{2}+\left(  X_{2}-b_{i}X_{0}\right)
^{2}\right]  \right\}  }{%
{\displaystyle\prod\limits_{i=1}^{N}}
\left[  \left(  X_{1}-a_{i}X_{0}\right)  ^{2}+\left(  X_{2}-b_{i}X_{0}\right)
^{2}\right]  }%
\end{array}
\right.  \label{8}%
\end{equation}

Let denominator through: $L=L\left(  \,X_{0},X_{1},X_{2}\right)  =%
{\displaystyle\prod\limits_{i=1}^{N}}
\left[  \left(  X_{1}-a_{i}X_{0}\right)  ^{2}+\left(  X_{2}-b_{i}X_{0}\right)
^{2}\right]  $.

Transforming further we have: $\left\{
\begin{array}
[c]{l}%
y_{1}=\frac{X_{1}}{X_{0}}-\frac{X_{0}\sum\limits_{j=1}^{N}m_{j}\left(
X_{1}-a_{j}X_{0}\right)  \left\{
{\displaystyle\prod\limits_{i=1,i\neq j}^{N}}
\left[  \left(  X_{1}-a_{i}X_{0}\right)  ^{2}+\left(  X_{2}-b_{i}X_{0}\right)
^{2}\right]  \right\}  }{L\left(  \,X_{0},X_{1},X_{2}\right)  }\\
y_{2}=\frac{X_{2}}{X_{0}}-\frac{X_{0}\sum\limits_{j=1}^{N}m_{j}\left(
X_{2}-b_{j}X_{0}\right)  \left\{
{\displaystyle\prod\limits_{i=1,i\neq j}^{N}}
\left[  \left(  X_{1}-a_{i}X_{0}\right)  ^{2}+\left(  X_{2}-b_{i}X_{0}\right)
^{2}\right]  \right\}  }{L\left(  \,X_{0},X_{1},X_{2}\right)  }%
\end{array}
\right.  $

We transform the equation to a polynomial form.%

\begin{equation}
\left\{
\begin{array}
[c]{l}%
\left(  X_{1}-y_{1}X_{0}\right)  L\left(  \,X_{0},X_{1},X_{2}\right)
-X_{0}^{2}\sum\limits_{j=1}^{N}m_{j}\left(  X_{1}-a_{j}X_{0}\right)  \left\{
{\displaystyle\prod\limits_{i=1,i\neq j}^{N}}
\left[  \left(  X_{1}-a_{i}X_{0}\right)  ^{2}+\left(  X_{2}-b_{i}X_{0}\right)
^{2}\right]  \right\}  =0\\
\left(  X_{2}-y_{2}X_{0}\right)  L\left(  \,X_{0},X_{1},X_{2}\right)
-X_{0}^{2}\sum\limits_{j=1}^{N}m_{j}\left(  X_{2}-b_{j}X_{0}\right)  \left\{
{\displaystyle\prod\limits_{i=1,i\neq j}^{N}}
\left[  \left(  X_{1}-a_{i}X_{0}\right)  ^{2}+\left(  X_{2}-b_{i}X_{0}\right)
^{2}\right]  \right\}  =0
\end{array}
\right.  \label{9}%
\end{equation}

Note that the polynomial in the left side of the first (second) equation is homogeneous.

For the degrees of these polynomials we have:

$\deg\left\{  \left(  X_{1}-y_{1}X_{0}\right)  L\left(  \,X_{0},X_{1}%
,X_{2}\right)  -X_{0}^{2}\sum\limits_{j=1}^{N}m_{j}\left(  X_{1}-a_{j}%
X_{0}\right)  \left\{
{\displaystyle\prod\limits_{i=1,i\neq j}^{N}}
\left[  \left(  X_{1}-a_{i}X_{0}\right)  ^{2}+\left(  X_{2}-b_{i}X_{0}\right)
^{2}\right]  \right\}  \right\}  =2N+1$, $\deg\left\{  \left(  X_{2}%
-y_{2}X_{0}\right)  L\left(  \,X_{0},X_{1},X_{2}\right)  -X_{0}^{2}%
\sum\limits_{j=1}^{N}m_{j}\left(  X_{2}-b_{j}X_{0}\right)  \left\{
{\displaystyle\prod\limits_{i=1,i\neq j}^{N}}
\left[  \left(  X_{1}-a_{i}X_{0}\right)  ^{2}+\left(  X_{2}-b_{i}X_{0}\right)
^{2}\right]  \right\}  \right\}  =2N+1.$

It's obvious that:%

\begin{equation}
\deg F_{1}=\deg F_{2}=2N+1 \label{10}%
\end{equation}

\subsection{Estimating the number of solutions of homogeneous polynomials}

Let the number of solutions of (\ref{7}), of course. The case in which the
number of solutions endlessly, it is found out quite simple, and does not need
to move to the projective coordinates, see. e.g., \cite{Kotv}.

To estimate the number of solutions of homogeneous polynomials (\ref{7}) we
apply the following B\'{e}zout's theorem, see. e.g., \cite{Reid}, which says:
that if $C$ and $D$ are plane curves of degrees $\deg C=m$, $\deg D=n$, then
the number of points of intersection of $C$ and $D$ is $m\cdot n$, provided
that \newline(i)\hspace{5mm} the field is algebraically closed;\newline%
(ii)\hspace{5mm} points of intersection are counted with the right
multiplicities;\newline(iii)\hspace{5mm} we work in $P^{2}$ to take right
account of intersections `at infinity'.

See for example [10].

Note that the equations $F_{1}\left(  \,X_{0},X_{1},X_{2},y1\right)  =0$ and
$F_{2}\left(  \,X_{0},X_{1},X_{2},y_{2}\right)  =0$ define two curves in the
projective space, and considered over an algebraically closed field of complex
numbers. Thus we are in the conditions of application of the B\'{e}zout's theorem.

\section{The conclusions}

- Number of complex solutions of equations (\ref{7}), taking into account the
multiplicity, in the projective plane $P^{2}$ equal $(2N+1)^{2}$; \newline-The
number of real solutions of the system of equations (\ref{7}), taking into
account the multiplicity equal to $(2N+1)^{2}-2n$ in the projective plane
$P^{2}$, where $n$- a natural number, that is less than or equal to $N$, that
is $n\leq N$. The last relation occurs because the systems of equations
(\ref{7}), although considered over the field of complex numbers, but have
real coefficients. The fact, that in such equations, complex roots come in pairs.

-number $(2N+1)^{2}-2n$- odd number. The number of real solutions of the
system of equations (\ref{7}), taking into account the multiplicity, in the
affine plane $A^{2}$, is also equal to this number, if the curves do not
intersect at infinity.

Note that each real solution of system has an obvious physical meaning: the
coordinates of the point - image plane of the lens. Our results, about the
oddness of the number of images are in good agreement with the theorem proved
in \cite{Zakh}, \cite{Sch}. At the same time, our approach is based on the
methods of algebraic geometry.

\begin{center}

\end{center}

\end{document}